\documentstyle[12pt,epsfig,twoside,fleqn,espcrc1]{article}

\newcommand{\AmS}{{\protect\the\textfont2
  A\kern-.1667em\lower.5ex\hbox{M}\kern-.125emS}}

\title{Electromagnetic probes of dense matter in heavy-ion
collisions\thanks{Work supported by the Department of Energy under grant No. 
DE-FG-88ER40388 and by the National Science Foundation under 
grant No. PHY-9509266}} 

\author{G.Q. Li\address{Department of Physics,  
        State University of New York at Stony Brook, \\ 
        Stony Brook, NY 11793, USA}, G.E. Brown$^{\rm a}$, and
C.M. Ko\address{Cyclotron Institute and Physics Department,
        Texas A\&M University, \\
        College Station, TX 77843, USA}}

\begin{document}

\maketitle

\begin{abstract}
Dilepton and photon production in heavy-ion collisions at 
SPS energies are studied in the relativistic transport 
model that incorporates self-consistently the change of hadron
masses in dense matter. It is found that the dilepton spectra in 
proton-nucleus reactions can be well described by the conventional
mechanism of Dalitz decay and direct vector meson decay. However, 
to provide a quantitative explanation of the observed dilepton spectra 
in central heavy-ion collisions requires contributions other than 
these direct decays and also various medium effects. 
Introducing a decrease of vector meson masses in hot dense medium, we find 
that these heavy-ion data can be satisfactorily explained. Furthermore,
the single photon spectra in our calculations with either free or in-medium 
meson masses do not exceed the upper bound deduced from the experiments by 
the WA80 Collaboration.

\end{abstract}

\section{INTRODUCTION}
 
The measurement of electromagnetic observables such as the photon 
and dilepton spectra constitutes a major part of CERN-SPS 
heavy-ion programs. It will soon become one of major efforts
at the BNL-RHIC heavy-ion programs. The primary reason for
this is that photons and dileptons do not suffer
strong final-state interactions as hadrons do. They can thus
be considered as `penetrating probes' of the initial hot and dense
stages involved in high-energy heavy-ion collisions.

Recent observation of the enhancement of low-mass dileptons in
central heavy-ion collisions at CERN-SPS energies 
\cite{ceres95,helios95} has generated a great deal of 
interest in the heavy-ion community. Different dynamical 
models, such as the hydrodynamical and transport models, 
have been used to investigate this phenomenon 
\cite{li95b,cass95a,gale95,wam95,koch96,steele96,hung96,soll96,red96,haglin96}. 
Calculations based on `conventional sources' such 
as the Dalitz decay and direct vector
meson decay that account for the dilepton spectra
in proton-induced reactions failed to explain the 
observed enhancement in heavy-ion collisions.
Various medium effects, such as the dropping vector meson masses
\cite{li95b,cass95a,hung96,red96} as first proposed by Brown and 
Rho \cite{br91}, the modification of rho meson spectral function 
\cite{wam95}, and the enhanced production of
$\eta$ and/or $\eta^\prime$ \cite{huang96}, have been proposed
to explain this enhancement. The last possibility has been ruled out by
the measured $\eta$ spectra in the same reactions \cite{drees96}.
The effects due to medium modifications of rho meson spectral function on
low-mass dileptons have been examined using simplified 
fireball models \cite{wam95}; questions remain 
concerning the exact space-time evolution and baryon 
chemical compositions which play an important role in the 
determination of the rho meson spectral function. 
In this contribution we will concentrate on the
idea of dropping vector meson masses, which is the focus
of Section 3. 

Another piece of experimental data from CERN-SPS that has been 
discussed extensively is the single photon spectra from the WA80 
collaboration \cite{wa80}. In various hydrodynamics calculations 
the absence of significant thermal photon production has been used as 
an indication of quark gluon plasma formation that lowers the 
initial temperature \cite{hydro}. However, detailed transport model 
calculations find this conclusion to be premature. 
This will be discussed in Section 4. A brief summary is given in
Section 5.

\section{THE RELATIVISTIC TRANSPORT MODEL}

In studying medium effects in heavy-ion collisions, the relativistic 
transport model \cite{ko87} based on the Walecka-type model \cite{qhd86}
has been quite useful, as it provides a thermodynamically consistent 
description of the medium effects through the scalar and vector
fields. In heavy-ion collisions at CERN-SPS energies, 
many hadrons are produced in the initial nucleon-nucleon 
interactions. This is usually modeled by the fragmentation of 
strings, which are the chromoelectric flux-tubes 
excited from the interacting quarks. One successful model 
for taking into account this nonequilibrium dynamics is the RQMD model 
\cite{sorge89}. To extend the relativistic transport model to heavy-ion 
collisions at these energies, we have used as initial conditions the 
hadron abundance and distributions obtained from the string fragmentation 
in RQMD.  Further interactions and decays of these hadrons are then 
taken into account as in usual relativistic transport model.

To study the effects of dropping vector meson masses 
\cite{br91,hatsuda92} on the dilepton spectrum in heavy-ion 
collisions, we have extended the Walecka model from the 
coupling of nucleons to scalar and vector fields to the 
coupling of light quarks to these fields, using the 
ideas of the meson-quark coupling model \cite{thomas94}.
For a system of nucleons, pseudoscalar mesons,
vector mesons, and axial-vector mesons
at temperature $T$ and baryon density $\rho _B$, the scalar field 
$\langle\sigma\rangle$ is determined self-consistently from 
\begin{eqnarray}
m_\sigma^2\langle \sigma\rangle &=&{4g_\sigma\over (2\pi )^3}\int d{\bf k} 
{m_N^*\over E^*_N}\Big[{1\over \exp ((E^*_N-\mu _B)/T)+1}
+{1\over \exp ((E^*_N+\mu _B)/T)+1}\Big]\nonumber\\
&+&{0.45g_\sigma\over (2\pi )^3}\int d{\bf k} {m_\eta^*\over E_\eta ^*}
{1\over \exp (E_\eta ^*/T)-1}+
{6g_\sigma\over (2\pi )^3}\int d{\bf k} {m_\rho^*\over E_\rho ^*}
{1\over \exp (E_\rho ^*/T)-1}\nonumber\\
&+&{2g_\sigma\over (2\pi )^3}\int d{\bf k} {m_\omega^*\over E_\omega ^*}
{1\over \exp (E_\omega ^*/T)-1}
+{6\sqrt 2 g_\sigma\over (2\pi )^3}\int d{\bf k} {m_{a_1}^*\over E_{a_1}^*}
{1\over \exp (E_{a_1}^*/T)-1},
\end{eqnarray}
where we have used the constituent quark model relations for 
the nucleon and vector meson masses \cite{thomas94}, 
i.e., $m_N^*=m_N-g_\sigma\langle \sigma\rangle ,
~m_{\rho ,\omega}^*\approx m_{\rho ,\omega}
-(2/3)g_\sigma\langle\sigma\rangle$, the quark 
structure of the $\eta$ meson in free space which leads to 
$m_\eta^*\approx m_\eta -0.45g_\sigma\langle\sigma\rangle$, and 
the Weinberg sum rule relation between the rho-meson and $a_1$ 
meson masses, 
i.e, $m_{a_1}^*\approx m_{a_1}-(2\sqrt 2/3)g_\sigma\langle\sigma\rangle$.
We recently found that the use of a refined model, the effective
chiral Lagrangian of \cite{fst},
led to essentially the same results for dilepton spectra \cite{li97}.

\begin{figure}[htb]
\begin{center}
\epsfig{file=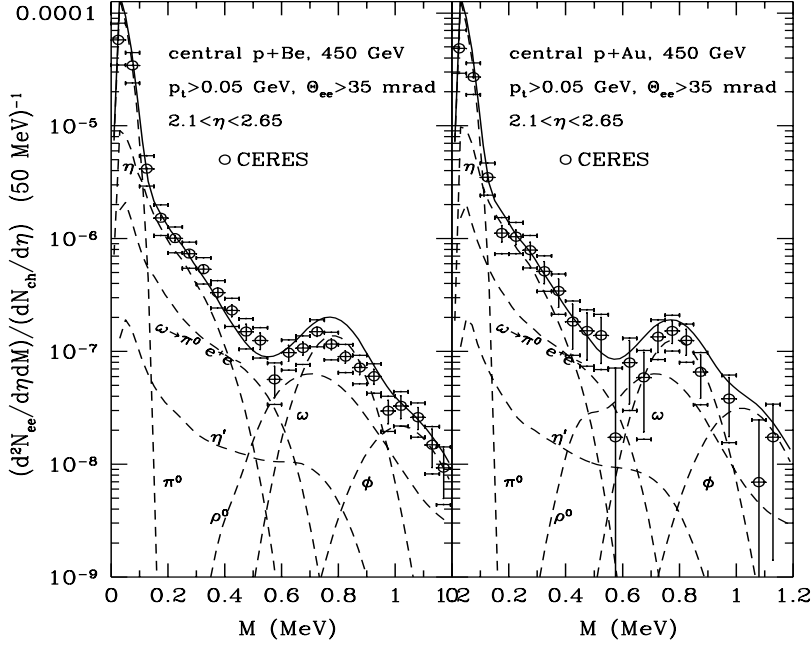,height=3.8in,width=4.5in}
\caption{Dilepton invariant mass spectra in p+Be (left window)
and p+Au (right window) collisions at 450 GeV.}
\end{center}
\end{figure}

\section{DILEPTON PRODUCTION}

The main contributions to dileptons with mass below 1.2 GeV are the
Dalitz decay of $\pi^0$, $\eta$ and $\omega$, the direct leptonic decay 
of $\rho^0$, $\omega$ and $\phi$, the pion-pion annihilation which 
proceeds through the $\rho^0$ meson, and the kaon-antikaon annihilation 
that proceeds through the $\phi$ meson.  The differential widths for 
the Dalitz decay of $\pi^0$, $\eta$, and $\omega$ are related to their 
radiative decay widths via the vector dominance model, which are taken 
from Ref. \cite{land85}.  

The decay of a vector meson into the dilepton is determined by the width,
\begin{equation}\label{lwidth}
\Gamma_{V\to l^+l^-}(M)= C_{l^+l^-}
\frac{m_V^4}{3M^3}(1-\frac{4m_l^2}{M^2})^{1/2}(1+\frac{2m_l^2}{M^2}).
\end{equation}
The coefficient 
$C_{l^+l^-}$ in the dielectron channel is $8.814\times 10^{-6}$,
$0.767\times 10^{-6}$, and $1.344\times 10^{-6}$ for $\rho$, $\omega$,
and $\phi$, respectively, and is determined from the measured width.
For the dimuon channel, these values are slightly larger.

In our model, dileptons are emitted continuously during the time evolution 
of the colliding system.  The way the dilepton yield is calculated can 
be illustrated by the decay of a rho meson. Denoting, at time $t$, the 
differential 
multiplicity of neutral rho mesons by $dN_{\rho ^0} (t)/dM$, then the 
differential dilepton production probability is given by
\begin{eqnarray}\label{sum}
{dN_{l^+l^-}\over dM} =\int _0^{t_f} {dN_{\rho^0} (t)\over dM} 
\Gamma _{\rho^0
\rightarrow l^+l^-}(M) dt + {dN_{\rho ^0} (t_f)\over dM} 
{\Gamma _{\rho^0\rightarrow l^+l^-} (M)\over\Gamma _\rho (M)},
\end{eqnarray}
where $t_f$ is the freeze-out time, which is found to be about 20 fm/c.
The first term corresponds to dilepton emission before freeze out while
the second term is from the decay of rho mesons still present at freeze out.
 
\begin{figure}[htb]
\begin{center}
\epsfig{file=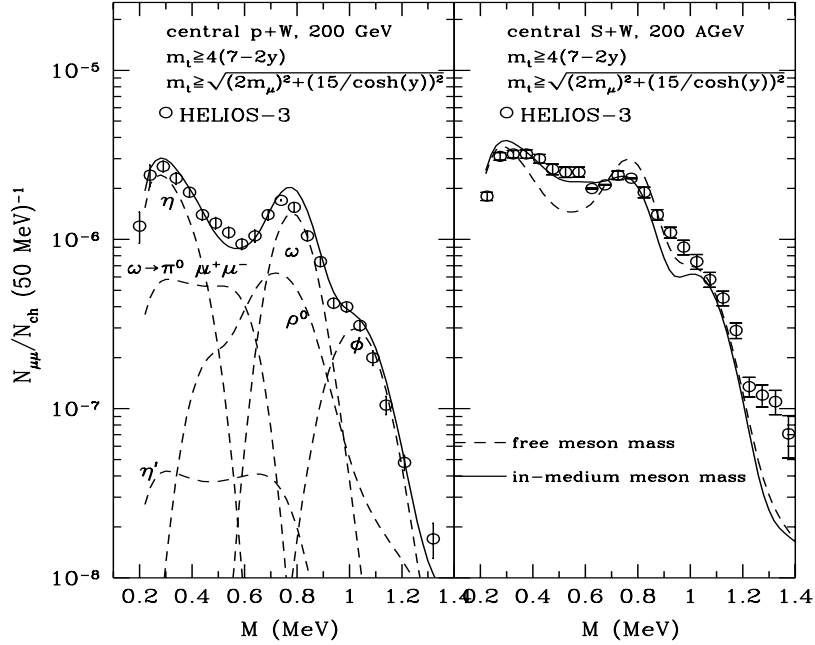,height=3.8in,width=4.5in}
\caption{Dilepton invariant mass spectra in p+W  
(left window) and S+W (right window) collisions at 200 AGeV. 
In the right window the solid and dashed curves are obtained with
in-medium and free meson masses, respectively.}
\end{center}
\end{figure}

The results for dilepton spectra from p+Be and p+Au collisions 
at 450 GeV are shown in Fig. 1, together with data from the 
CERES \cite{ceres95}. It is seen that the data can be well 
reproduced by Dalitz decay of $\pi^0$, $\eta$ and $\omega$ 
mesons, and direct leptonic decay of $\rho^0$, $\omega$ 
and $\phi$ mesons. These results are thus similar 
to those found in Ref. \cite{cass95a} using the Hadron-String Dynamics
and those constructed by the CERES collaboration from known and expected 
sources of dileptons \cite{ceres95}. Similar conclusions can
be drawn concerning dimuon spectra in p+W collisions at
200 GeV, which are shown in the left window of Fig. 2.
 
Our results for dilepton spectra in central S+Au collisions
are shown in the left window of Fig. 3. The dashed 
curve is obtained with free meson masses. Although 
pion-pion annihilation is important for dileptons with 
invariant mass from 0.3 to 0.65 GeV, it still does not 
give enough number of dileptons in this mass region.
Furthermore, for masses around $m_{\rho,\omega}$ there are 
more dileptons predicted by the theoretical calculations than shown in 
the experimental data. These are very similar to the results 
of Cassing {\it et al.} \cite{cass95a} based on the 
Hadron-String Dynamics model and Srivastava {\it et al.} \cite{gale95}
based on the hydrodynamical model.
The results obtained with in-medium meson masses are shown 
by the solid curve.  Compared with the results obtained with free meson 
masses, there is about a factor of 2-3 enhancement of the dilepton yield 
in the mass region from 0.2 to 0.6 GeV, which thus leads to a good 
agreement between the theoretical results and the CERES data. 
Similar conclusions that dropping vector
meson masses can explain the CERES dilepton data have been
obtained in Refs. \cite{cass95a,hung96,red96}. 
 
\begin{figure}[htb]
\begin{center}
\epsfig{file=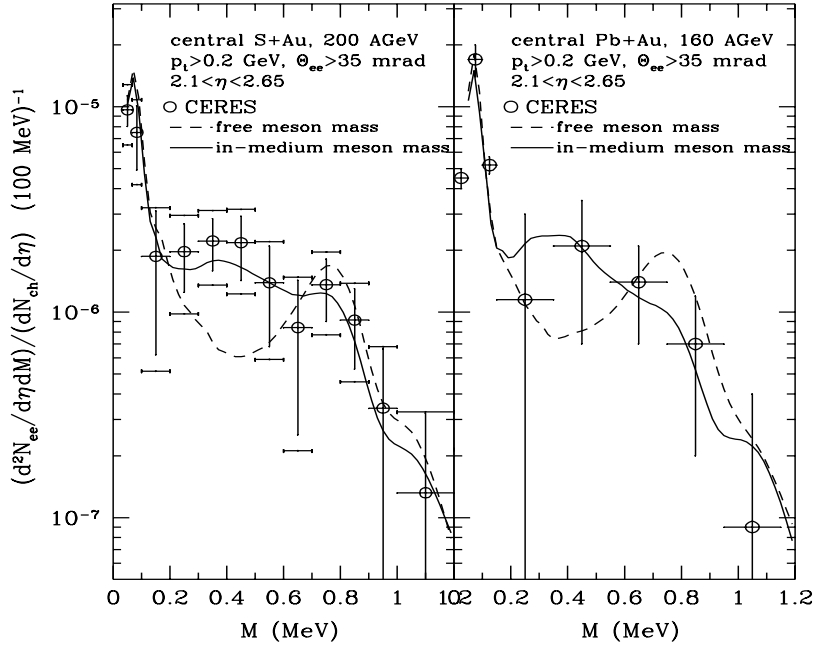,height=3.8in,width=4.5in}
\caption{Dilepton invariant mass spectra in S+Au collision at 
200 AGeV (left window) and Pb+Au collisions at 160 AGeV 
(right window). The solid and dashed curves are obtained with
in-medium and free meson masses, respectively.}
\end{center}
\end{figure}

In the right window of Fig. 3 we compare our prediction for
central Au+Pb collisions at 160 AGeV with the preliminary
data from the CERES collaboration. 
The normalization factor $dN_{ch}/d\eta$ here is the average charge 
particle pseudo-rapidity density in the pseudo-rapidity range of 2 
to 3, and is about 440 in this collision. In the results with free meson 
masses, shown by the dashed curve, there 
is a strong peak around $m_{\rho ,\omega}$, which is dominated by 
$\rho^0$ meson decay as a result of an enhanced contribution from 
pion-pion annihilation in Pb+Au collisions than in S+Au and 
proton-nucleus collisions. In the case of in-medium meson masses, 
shown by the solid curve, the $\rho$ 
meson peak shifts to a lower mass, and the peak around $m_{\rho ,\omega}$ 
becomes a shoulder arising mainly from $\omega$ meson decay. At the same 
time we see an enhancement of low-mass dileptons in the region of 
0.25-0.6 GeV as in S+Au collisions. The agreement with the data
is thus significantly improved when dropping vector meson masses are used.      

The same model has been used to calculate the dimuon spectra 
from central S+W collisions. The results obtained
with free meson masses are shown in the right window of Fig. 2
by the dashed curve, and are below the HELIOS-3 data in the mass 
region from 0.35 to 0.6 GeV, and slightly above the data 
around $m_{\rho ,\omega}$ as in the CERES case. However, 
the discrepancy between the theory and the data is somewhat 
smaller in this case due to the smaller charged-particle multiplicity at 
a larger rapidity than in the CERES experiment.
Our results obtained with in-medium meson masses are shown in 
the right window of Fig. 2 by the solid curve, and are 
in good agreement with the data.  The importance of 
dropping rho meson mass in explaining the HELIOS-3 data has also been 
found by Cassing {\it et al.} \cite{cass95a}.
 
\section{PHOTON PRODUCTION}

Single photon spectra in heavy-ion collisions at CERN-SPS energies
have been measured by the WA80 collaboration \cite{wa80}. So
far, only the upper bound has been determined for the so-called
`thermal' photon spectra, which accounts for about 5\% of the
total observed single photon yield, which is dominated by neutral
pion decay. In hydrodynamical calculations \cite{hydro}, the absence
of significant thermal photons has been interpreted as an
evidence for the formation of a quark gluon plasma. Without
phase transition, the initial temperature of the hadronic gas
found in Ref. \cite{hydro} is about 400 MeV. This leads to
a large number of thermal photons from hadronic interaction which
is not observed experimentally. Including the phase transition,
the initial temperature is lowered to about 200 MeV \cite{hydro},
because of increased degrees of freedom. With a lower initial
temperature, the thermal photon yield is thus reduced and is found
to better agree with the WA80 data. However, a more recent analysis 
\cite{cley96} has shown that if one includes all hadron resonances with 
masses below 2.5 GeV the initial temperature can also be lowered to 
about 200 MeV, and the WA80 photon data can then be explained without 
invoking the formation of a quark gluon plasma. This is basically also the
conclusion of the detailed transport model analysis presented here.

For the `thermal' photon spectra we include the decay of $\rho$, $\omega$,
$\eta^\prime$, and $a_1$ mesons, as well as two-body processes such
as $\pi\pi\rightarrow \rho \gamma$ and $\pi\rho\rightarrow \pi\gamma$.
The decay width for $\rho\rightarrow \pi\pi\gamma$ is taken from the model
of Ref. \cite{sig63}, which describes well the measured width for 
$\rho^0\rightarrow \pi\pi\gamma$. Both the $\omega$ and $a_1$ radiative 
decay widths are proportional to $|{\bf p}_\pi |^3$ \cite{shur91}, with the
coefficients determined from the measured width. For the two-body
cross sections, we use the results of \cite{kapu91}, which do not
include the contribution from an intermediate $a_1$ meson. 
The latter has already been included in our model as a two-step process. 

In Fig. 4, our results for `thermal' photon spectra are compared with the
upper bound from the WA80 collaboration.
It is seen that the $\omega$ meson radiative decay plays an
important role for photons with transverse momenta above
300 MeV. In both the free and in-medium meson mass scenarios,
our thermal photon yields do not exceed the experimental upper bound.
Although there are more rho and $a_1$ mesons in the case of in-medium
meson masses, their contributions to single photon yield do not increase
significantly, simply because their radiative decay widths
decrease as a result of reduced phase space.

\begin{figure}[htb]
\begin{center}
\epsfig{file=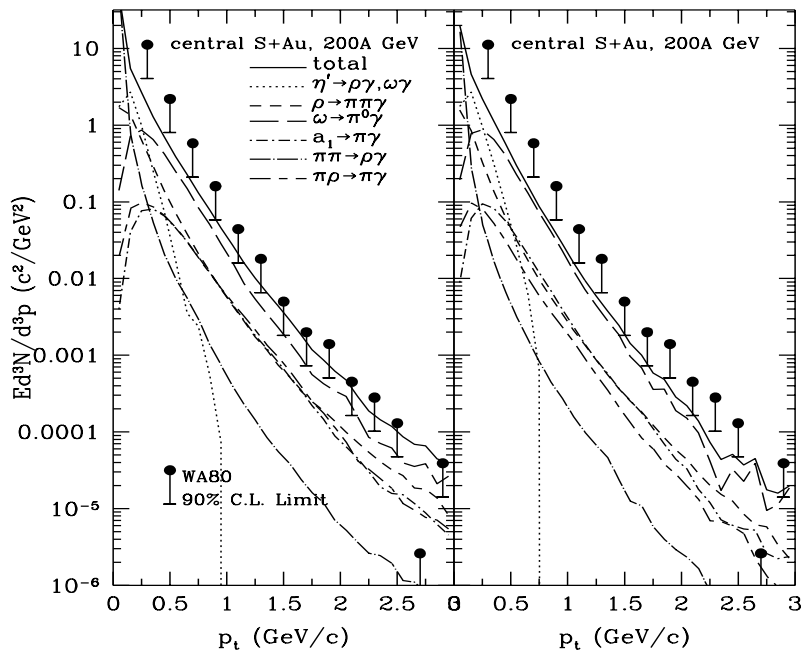,height=3.8in,width=4.5in}
\caption{`Thermal' photon spectra in central S+Au collisions at 
200 AGeV using free (left window) and in-medium 
(right window) meson masses.}
\end{center}
\end{figure}

\section{SUMMARY AND OUTLOOK}
 
In summary, we have studied dilepton production from both 
proton-nucleus and nucleus-nucleus collisions using the 
relativistic transport model with initial conditions 
determined by string fragmentation from the initial stage 
of the RQMD model.  It is found that the dilepton spectra 
in proton-nucleus reactions measured by the CERES and the HELIOS-3 
collaboration can be well understood in terms of conventional 
mechanisms of Dalitz decay and direct vector meson decay. 
For dilepton spectra in central heavy-ion collisions, these 
conventional mechanisms, however, fail to explain the data, 
especially in the low-mass region from about 0.25 to 
about 0.6 GeV in CERES experiments, and from 0.35 to 0.65 GeV 
in HELIOS-3 experiments. Including the contribution from pion-pion 
annihilation, which is important in the mass region from 
$2m_\pi$ to $m_{\rho ,\omega}$, removes some of the discrepancy.  
But the theoretical prediction is still substantially below the 
data in the low mass region and somewhat above the data around 
$m_{\rho,\omega}$. The theoretical results are brought into good 
agreement with the data when reduced in-medium vector meson masses 
are taken into account. We have also calculated, within 
the same dynamical model, the thermal photon spectra in 
central S+Au collisions. In both free meson masses and 
in-medium meson masses scenarios, our results do not exceed 
the upper bound deduced from the experiments by the WA80 collaboration.

\end{document}